%Paper: hep-ph/9210261
%From: rhb@het.brown.edu (Robert Brandenberger)
%Date: Sat, 24 Oct 92 16:01:59 EDT

\input phyzzx
\FRONTPAGE
\line{\hfill BROWN-HET-881}
\line{\hfill September 1992}
\vskip1.5truein
\titlestyle{{TOPOLOGICAL DEFECT MODELS OF STRUCTURE FORMATION \break
AFTER THE COBE DISCOVERY OF CMB ANISOTROPIES}\foot{Invited lecture at
the International School of Astrophysics ``D. Chalonge," 2nd course,
6-13 Sept., 1992, Erice, Italy, to be published in the proceedings
(World Scientific, Singapore, 1992).}\break}
\bigskip
\author{Robert H. BRANDENBERGER}
\centerline{{\it Department of Physics}}
\centerline{{\it Brown University, Providence, RI 02912, USA}}
\bigskip
\abstract
The COBE satellite has discovered anisotropies in the cosmic microwave
background (CMB) consistent with a scale invariant spectrum of density
perturbations$^{1)}$. As reviewed in this lecture, topological defect models of
structure formation generically produce such a spectrum.
We give a summary of the cosmic string and global texture models, focusing on
distinct observational
signatures in large-scale structure surveys and in the microwave background.
It is shown that the amplitude of the CMB quadrupole
detected by COBE is in good agreement with the predictions of the
cosmic string model.$^{2,3)}$
\endpage
\chapter{Introduction}
\par
One aim of this review is to compare the three most popular models of
structure formation in the light of the new COBE results$^{1)}$ on
anisotropies in the CMB.  These models are:
\item{-} The ``CDM model,"$^{4)}$ a theory based on quantum
fluctuations from inflation
as the seeds of structures and dark matter
being cold.
\item{-} The cosmic string theory with hot dark matter.$^{5)}$
\item{-} The global texture model with cold dark matter.$^{6)}$

\noindent
Contrary to what is often stated, the recent COBE results do not
provide evidence for inflation.

In order to demonstrate the above point, a review of topological
defect models of structure formation will be presented, emphasizing
the reasons why all such models give a scale invariant spectrum of
density perturbations.

The third goal of this lecture is to discuss some good statistics for
distinguishing between the three models.  Such statistics must be
sensitive to the nonrandom phases of the density field of topological
defect theories.  Root mean square measures ({\it e.g.} the CMB
quadrupole) do not satisfy this criterion.

The outline of this article is as follows:  First, I will briefly
compare the three models of structure formation listed above in the
light of the recent COBE results.  Next, I will give a short review of
cosmic strings and textures, focusing on the key differences.  Section
4 is an outline of the cosmic string and global texture models of
structure formation.  In Section 5 I discuss specific signatures for
strings and textures and focus on statistics which can distinguish
between models with the same spectrum of density perturbations but
different phases.

Throughout this article, units in which $k_B = \hbar = c = 1$ are
used.  Unless otherwise indicated, the Hubble expansion rate $H$ will
be taken to be 50 kms$^{-1}$ Mpc$^{-1}$.  Calculations assume a
spatially flat Robertson-Walker metric with scale factor $a(t)$, and
$z(t)$ denotes the redshift at time $t$.  $G$ is Newton's constant,
and $m_{p\ell}$ is the Planck mass.

\chapter{A First Comparison and the Connection with COBE Results}

The two classes of models of structure formation which have been
studied in greatest depth over the past few years are
\item{i)} quantum fluctuations from inflation
\item{ii)} topological defects \hfill\break
a) cosmic strings$^{7)}$ \hfill\break
b) global monopoles$^{8,9)}$ \hfill\break
c) global textures$^{10,6)}$

\noindent
This list is not exhaustive.  For example, models based on late time
phase transitions$^{11)}$ are not included.

In terms of their predictions there are important similarities but
also crucial differences between the two classes of models.  Both give
a scale invariant spectrum of density perturbations
$$
\delta (k) \sim k^n \, , \> n = 1 \, , \eqno\eq
$$
where $\delta (k)$ is the Fourier transform of the fractional density
contrast ${{\delta \rho} \over {\rho}} (\undertext{x})$.  Equation (2.1) is
valid
on scales larger than the comoving Hubble radius at $t_{eq}$, the time
of equal matter and radiation.

However, as regards to the coherence of the phases of perturbations of
different wavelengths, there are crucial differences (see Fig. 1).
Quantum fluctuations from inflation produce random phase
superpositions of fluctuations on all wavelengths (see the lecture by
L. Grishchuk and Refs. 12 and 13 for some subtle issues), whereas
topological defects give rise to nonrandom phases.

\midinsert \vskip 4.0cm
\hsize=6in \raggedbottom
\noindent{\bf Figure 1.}
A comparison between spectra of perturbations with random (left) and
nonrandom phases.
\endinsert

Both classes of models have free parameters.  In inflationary
Universe models (to be specific we shall consider chaotic inflation
driven by a scalar field $\varphi$ with potential
$$
V (\varphi) = {1\over 4} \lambda \varphi^4 )\, , \eqno\eq
$$
the coupling constant $\lambda$ is usually a free parameter.  To
obtain the right order of magnitude for density perturbations, a value
$\lambda \sim 10^{-12}$ is required.$^{14)}$  However, if $\varphi$ is
coupled to the gauge fields of the standard model (of particle
physics) or of grand unified theories, a value of $\lambda \sim 1$ is
induced naturally by quantum effects.  This discrepancy is a serious
problem for inflationary models of structure formation.

The free parameter in topological defect theories is the scale $\eta$
of symmetry breaking.  For example, the mass per unit length $\mu$ of
a cosmic string is $\mu \simeq \eta^2$.  In order to match large-scale
structure observations, a value $\eta \sim 10^{16}$ GeV is
required.$^{15, 5)}$  In encouraging agreement between astronomy and
particle physics, this scale is the scale of symmetry breaking in many
grand unified models.

The amplitude of the power spectrum in all of the theories considered
here can -- after the recent COBE discovery of anisotropies in the CMB
-- be determined in various ways.  The first possibility is to use
large-scale structure data (either streaming motion or clustering
strength) to set the amplitude.  I shall denote the amplitude
determined this way by $A_1$.  Alternatively, the magnitude of CMB
anisotropies can be used to determine the amplitude.  The result of
this procedure is $A_2$.

Note that the COBE results published to date$^{1)}$ only give the CMB
quadrupole and a few low harmonies.  The data is consistent with a
scale invariant spectrum.  The quadrupole temperature anisotropy is
$$
{\delta T\over T} \Big|_q \sim 6 \cdot 10^{-6} \, . \eqno\eq
$$
\par
It should be obvious that given the present results, COBE
\undertext{cannot} distinguish between our two classes of models.
Pixel by pixel information is required in order to be able to
calculate statistics which allow one to differentiate between random
phase and the different nonrandom phase models.  At present, the only
published limit on pixel signals is$^{1)}$
$$
{\delta T\over T} \Big|_{\scriptstyle{\rm pixel}} < 8 \cdot 10^{-
5} \, . \eqno\eq
$$
\par
However, for every individual theory we may ask whether the different
normalizations of the amplitude of the power spectrum give the same
result.  For the inflationary CDM model the amplitudes only agree if
the bias parameter $b \simeq 1$.$^{16)}$  Here, the bias parameter
gives the excess of light clustering over mass clustering
$$
{\delta L\over L} = b \, {\delta M\over M} \eqno\eq
$$
on a scale of 8Mpc (where $L$ stands for luminosity).  However, in
order to get sufficient large-scale structure relative to the smaller
scale structure, a bias parameter of the order $b \simeq 2.5$ is
required.$^{17)}$  In other words, the CMB quadrupole is further
evidence that the CDM model has insufficient power on large scales (or
equivalently too much power on small scales if we normalize according
to COBE).

For cosmic strings, recent numerical work by Bennett, Bouchet and
Stebbins$^{2)}$ and analytical work by Perivolaropoulos$^{3)}$ has
shown that the COBE and large-scale structure normalizations of the
model agree well.

For textures, there is a mismatch in normalization by a factor of
between 2.5 and 4$^{18-20)}$ (the COBE normalization is lower).
Biasing might be invoked to explain the difference.  However, it is
also important to note that the COBE normalization depends on the
specific form of the scaling solution, which (as in the case of cosmic
strings) is not yet accurately determined.

As a preliminary summary, we stress that after COBE topological defect
models are well and alive.  In the case of cosmic strings, there is up
to this point excellent agreement.  Thus, there is good justification
to turn to a review of topological defect models.

\chapter{Cosmic Strings and Global Textures}
\par
Consider a theory in which matter consists of a gauge field $A_\mu$ and a
complex scalar field $\phi$ whose dynamics is given by the Lagrangean
$$
{\cal L} = {1\over 2} \, D_\mu \phi D^\mu \phi - V (\phi) + {1\over 4} \,
F_{\mu \nu} \, F^{\mu \nu}  \eqno\eq
$$
where $D_\mu = \partial_\mu + ie\, A_\mu$
is the gauge-covariant derivative, $e$ is
the gauge coupling constant and $F_{\mu \nu}$ is the field strength tensor.
The potential $V(\phi)$ has the symmetry breaking ``Mexican hat" shape (see
Figure 2):
$$
V (\phi) = {1\over 4} \lambda (|\phi|^2 - \eta^2)^2 \, . \eqno\eq
$$
Hence, the vacuum manifold ${\cal M}$, the space of minimum energy density
configurations, is a circle $S^1$.
\midinsert \vskip 6.5cm
\hsize=6in \raggedbottom
\noindent{\bf Figure 2.} The zero temperature potential energy of the
complex scalar field used in the cosmic string model.
\endinsert
\par
The theory described by (3.1) and (3.2) admits one dimensional topological
defects, cosmic strings$^{21)}$.  It is possible to construct string
configurations which are translationally invariant along the $z$ axis.  On a
circle $C$ in the $x-y$ plane with radius $r$, the boundary conditions
for $\phi$ are
$$
\phi (r, \theta) = \eta \, e^{i \theta} \, \eqno\eq
$$
where $\theta$ is the polar angle along $C$. The
configuration (3.3) has winding number 1: at all points of the circle, $\phi$
takes on values in ${\cal M}$, and as $\theta$ varies from 0 to $2 \pi$,
$\phi$ winds once round ${\cal M}$.  By continuity of $\phi$ and since the
winding number is an integer, it follows that there must
be a point $p$ on the disk $D$ bounded by $C$ where $\phi = 0$.  By
translational symmetry
there is a line of points with $\phi = 0$.  This line is the center of the
cosmic string.  The cosmic string is a line of trapped potential energy.  In
order to minimize the total energy given the prescribed topology (i.e.
winding number), the thickness of the string (i.e. radius over which $V (\phi)$
deviates significantly from 0) must be finite.  As first shown in Ref.
22, the
width $w$ of a string is
$$
w \simeq \lambda^{-1/2} \eta^{-1} \, , \eqno\eq
$$
from which it follows that the mass per unit length $\mu$ is
$$
\mu \simeq \eta^2 \, . \eqno\eq
$$
\par
Cosmic strings arise in any model in which the vacuum manifold satisfies the
topological criterion
$$
\Pi_1 ({\cal M}) \neq {\bf 1} \, , \eqno\eq
$$
$\Pi_1$ being the first homotopy group.  Any field configuration $\phi
(\undertext{x})$ is characterized by an integer $n$, the element of $\Pi_1
({\cal M})$ corresponding to $\phi (\undertext{x})$.
\par
A cosmic string is an example of a \undertext{topological defect}.  A
topological defect has a well-defined core, a region in space where $\phi
\not\in {\cal M}$ and hence $V (\phi) > 0$.  There is an associated
winding number, and it is quantized.  Hence, a topological defect is stable.
Furthermore, topological defects exist for theories with global and local
symmetry groups.
\par
Cosmic strings are not the only topological defects.  In theories for which
the vacuum manifold ${\cal M}$ obeys $\Pi_0 ({\cal M}) \neq {\bf 1}$, two
dimensional defects -- domain walls -- exist.  An example is the theory of a
single real scalar field with symmetry breaking potential (3.2).  If the
theory contains three real scalar fields $\phi_i$ with potential (3.2) (if
$|\phi|^2 = \sum\limits^3_{i=1} \, \phi^2_i$), then $\Pi_2 ({\cal M})
\neq
{\bf 1}$ and monopoles result.
\par
Next, consider a theory of four real scalar fields given by the Lagrangean
$$
{\cal L} = {1\over 2} \partial_\mu \phi \partial^\mu \phi - V (\phi) \eqno\eq
$$
with
$$
V (\phi) = {1\over 4} \lambda \, \left( \sum\limits^4_{i=1} \, \phi^2_i -
\eta^2 \right)^2 \, . \eqno\eq
$$
In this case, the vacuum manifold ${\cal M} = S^3$ with topology
$$
\Pi_3 ({\cal M}) \neq {\bf 1} \, , \eqno\eq
$$
and the corresponding defects are the global textures$^{21,10,6)}$.
\midinsert \vskip 10cm
\hsize=6in \raggedbottom
\noindent{\bf Figure 3.} Construction of a global texture: left is
physical space, right the vacuum manifold. The field configuration
$\phi$ is a map from space to the vacuum manifold (see text).
\endinsert
\par
Textures, however, are quite different than the previous topological defects.
The texture construction will render this manifest (Fig. 3).  To construct a
radially symmetric texture, we give a field configuration $\phi (x)$ which
maps space onto ${\cal M}$.  The origin 0 in space (an arbitrary point which
will be the center of the texture) is mapped onto the north pole $N$ of ${\cal
M}$.  Spheres surrounding 0 are mapped onto spheres surrounding $N$.  In
particular, some sphere with radius $r_c (t)$ is mapped onto the equator
sphere of ${\cal M}$.  The distance $r_c (t)$ can be defined as the radius of
the texture.  Inside this sphere, $\phi (x)$ covers half the vacuum manifold.
Finally, the sphere at infinity is mapped onto the south pole of ${\cal M}$.
The configuration $\phi (\undertext{x})$ can be parameterized by$^{6)}$
$$
\phi (x,y,z) = \left(\cos \chi (r), \> \sin \chi (r) {x\over r}, \>
\sin \chi (r) {y\over r}, \> \sin \chi (r) {z\over r} \right) \eqno\eq
$$
in terms of a function $\chi (r)$ with $\chi (0) = 0$ and $\chi (\infty) =
\pi$.  Note that at all points in space, $\phi (\undertext{x})$ lies in ${\cal
M}$.  There is no defect core.  All the energy is spatial gradient (and
possibly kinetic) energy.
\par
In a cosmological context, there is infinite energy available in an infinite
space.  Hence, it is not necessary that $\chi (r) \rightarrow \pi$ as $r
\rightarrow \infty$.  We can have
$$
\chi (r) \rightarrow \chi_{\rm max} < \pi \>\> {\rm as} \>\> r \rightarrow
\infty \, . \eqno\eq
$$
In this case, only a fraction
$$
n = {\chi_{\rm max}\over \pi} - {\sin 2 \chi_{\rm max}\over{2 \pi}} \eqno\eq
$$
of the vacuum manifold is covered:  the winding number $n$ is not quantized.
This is a reflection of the fact that whereas topologically nontrivial maps
from $S^3$ to $S^3$ exist, all maps from $R^3$ to $S^3$ can be deformed to
the trivial map.
\par
Textures in $R^3$ are unstable.  For the configuration described above, the
instability means that $r_c (t) \rightarrow 0$ as $t$ increases: the texture
collapses.  When $r_c (t)$ is microscopical, there will be sufficient energy
inside the core to cause $\phi (0)$ to leave ${\cal M}$, pass through 0 and
equilibrate at $\chi (0) = \pi$: the texture unwinds.
\par
A further difference compared to topological defects: textures are relevant
only for theories with global symmetry.  Since all the energy is in spatial
gradients, for a local theory the gauge fields can reorient themselves such as
to cancel the energy:
$$
D_\mu \phi = 0 \, . \eqno\eq
$$
\par
Therefore, it is reasonable to regard textures as an example of a new class of
defects, \undertext{semitopological defects}.  In contrast to topological
defects, there is no core, and $\phi (\undertext{x}) \epsilon {\cal M}$ for all
$\undertext{x}$.  In particular, there is no potential energy.  Second, the
winding number is not quantized, and hence the defects are unstable.  Finally,
they exist only in theories with a global interval symmetry.
\par
The Kibble mechanism$^{21)}$ ensures that in theories which admit defects, such
defects inevitably will be produced during the symmetry breaking phase
transition in the very early Universe.  Causality implies that on scales
larger than the horizon the orientation of $\phi (\undertext{x})$ in the vacuum
manifold ${\cal M}$ will be random.  Hence, there is a finite probability $p$
per horizon volume that a defect will form (for a recent calculation of $p$
for various types of defects see Ref. 23).
\par
Applied to cosmic strings, the Kibble mechanism implies that at the time of
the phase transition $t_c$, a network of strings with mean separation $\xi
(t_c) < t_c$ will form.  The dynamical evolution of the string network is
complicated.  Causality alone implies that $\xi (t) < p^{-1/3} t$ for all $t$,
whereas a combined use of the Nambu action for string dynamics, the
intercommutation property$^{24)}$ of strings, and decay of loops by
gravitational radiation$^{25)}$ implies that $\xi (t) \sim t$ for the network
of infinite strings.  This was first argued heuristically in Ref. 7 (see Ref.
26 for general reviews), and was derived later by numerical
simulations$^{27)}$.
\par
Applied to textures (for a recent review see Ref. 28), the Kibble mechanism
implies that at all times $t > t_c$, the field configuration will be random on
scales $\xi > t$.  There is a probability $p (n)$$^{29)}$ that the field
configuration will be a texture with winding number $n > n_c$, where $n_c$ is
the critical winding$^{30)}$ above which a configuration collapses.
When such a texture configuration enters
the horizon, it will start to collapse.  Textures of radius $\sim t$
are thus ``created" at all
times $t > t_c$ with the same statistical correlations and distribution of
winding numbers.  This is the ``scaling solution" for textures.

To conclude this section I will demonstrate that the scaling solution
for strings and textures leads to a scale invariant spectrum.

The scaling solution implies that the r.m.s. mass perturbation $\delta
M/M \, (k, t)$ on a length scale $k^{-1}$ when measured at the time $t
= t_H (k)$ when this scale is equal to the Hubble radius is
independent of $k$:
$$
{\delta M\over M} \, (k, t_H (k)) = {\rm const} \, . \eqno\eq
$$
For example, in the cosmic string model $\delta M$ is the mass in
strings inside the Hubble radius, and $M$ is the total mass in this
volume.  Obviously, $\delta M \sim \mu t$ and $M \sim t^3 \rho (t)
\sim t$ (both in the matter and radiation dominated periods), and
hence (3.14) follows.

To convert (3.14) into a formula for the power spectrum (see {\it
e.g.} Ref. 31 for a survey of the relevant notation), we use the known
growth rate of perturbations in the matter dominated epoch to obtain
the mass spectrum at constant time $t$
$$
{\delta M\over M} \, (k, t) = \left( {t\over{t_H (k)}} \right)^{2/3}
\, {\delta M\over M} \, (k, t_H (k) ) \, . \eqno\eq
$$
For scales larger than the Hubble radius at $t_{eq}$, the time of
equal matter and radiation,
$$
t_H (k) = k^{-1} a (t_H (k)) \sim t_H^{2/3} (k) k^{-1} \, . \eqno\eq
$$
Hence
$$
t_H (k) \sim k^{-3} \eqno\eq
$$
and
$$
{\delta M\over M} \, (k, t) \sim k^{-2} \, . \eqno\eq
$$
Since the power spectrum $P(k)$ is related to $\delta M/M$ via
$$
({\delta M\over M})^2 (k, t) \simeq k^3 P(k) \, , \eqno\eq
$$
we conclude that
$$
P (k) \sim k \, , \eqno\eq
$$
{\it i.e.} the power spectrum has index $n = 1$ (scale invariant).

Therefore, as advertised earlier, both cosmic strings and textures
predict the same shape of the power spectrum as inflationary models,
and thus r.m.s. measurements cannot distinguish between them.

\chapter{Structure Formation with Cosmic Strings and Textures}
\par
The cosmic string model admits three mechanisms for structure
formation:  loops, filaments and wakes.  Cosmic string loops have the same time
averaged
field as a point source with mass
$$
M (R) = \beta R \mu \, , \eqno\eq
$$
$R$ being the loop radius and $\beta \sim 2 \pi$.  Hence, loops will be seeds
for spherical accretion of dust and radiation.
\midinsert \vskip 3.5cm
\hsize=6in \raggedbottom
\noindent{\bf Figure 4.} Sketch of the mechanism by which a long
straight cosmic string moving with velocity $v$ in transverse
direction through a plasma induces a velocity perturbations $\Delta v$
towards the wake. Shown on the left is the deficit angle, in the
center is a sketch of the string moving in the plasma, and on the
right is the sketch of how the plasma moves in the frame in which the
string is at rest.
\endinsert
\par
Long strings moving with relativistic speed in their normal plane give rise to
velocity perturbations in their wake.  The mechanism is illustrated in Fig. 4:
space normal to the string is a cone with deficit angle$^{32)}$
$$
\alpha = 8 \pi G \mu \, . \eqno\eq
$$
If the string is moving with normal velocity $v$ through a bath of dark
matter, a velocity perturbation
$$
\Delta v = 4 \pi G \mu v \gamma \eqno\eq
$$
[with $\gamma = (1 - v^2)^{-1/2}$] towards the plane behind the string
results.  At times after $t_{eq}$, this induces planar overdensities, the most
prominent (i.e. thickest at the present time) and numerous of which were
created at $t_{eq}$, the time of equal matter and radiation.$^{33, 34)}$  The
corresponding planar dimensions are (in comoving coordinates)
$$
t_{eq} z (t_{eq}) \times t_{eq} z (t_{eq}) v \sim (40 \times 40 v) {\rm Mpc}^2
\, . \eqno\eq
$$

The thickness $d$ of these wakes can be calculated using the
Zel'dovich approximation$^{35)}$.  The result is
$$
d \simeq G \mu v \gamma (v) z (t_{eq})^2 \, t_{eq} \simeq 4 v {\rm
Mpc} \, . \eqno\eq
$$
\par
Wakes arise if there is little small scale structure on the string.
In this case, the string tension equals the mass density, the string
moves at relativistic speeds, and there is no local gravitational
attraction towards the string.

In contrast, if there is small scale structure on strings,
then$^{36)}$ the string tension $T$ is smaller than the mass per unit
length $\mu$ and the metric of a string in $z$ direction becomes
$$
ds^2 = (1 + h_{00}) (dt^2 - dz^2 - dr^2 - (1 - 8G \mu) r^2 d\varphi^2 )
\eqno\eq
$$
with
$$
h_{00} = 4G (\mu - T) \ln \, {r\over r_0} \, , \eqno\eq
$$
$r_0$ being the string width.  Since $h_{00}$ does not vanish, there
is a gravitational force towards the string which gives rise to
cylindrical accretion, thus producing filaments.

As is evident from the last term in the metric (4.6), space
perpendicular to the string remains conical, with deficit angle given
by (4.2).  However, since the string is no longer relativistic, the
transverse velocities $v$ of the string network are expected to be
smaller, and hence the induced wakes will be shorter and thinner.

Which of the mechanisms -- filaments or wakes -- dominates is
determined by the competition between the velocity induced by $h_{00}$
and the velocity perturbation of the wake.  The total velocity is$^{37)}$
$$
u = - {2 \pi G (\mu - T)\over{v \gamma (v)}} - 4 \pi G \mu v \gamma
(v) \, , \eqno\eq
$$
the first term giving filaments, the second producing wakes.  Hence,
for small $v$ for former will dominate, for large $V$ the latter.

By the same argument as for wakes, the most numerous and prominent
filaments will have the distinguished scale
$$
t_{eq} z (t_{eq}) \times d_f \times d_f \eqno\eq
$$
where $d_f$ can be calculated using the Zel'dovich approximation.

In the texture model it is the contraction of the field configuration which
leads to density perturbations.  At the time when the texture enters the
horizon, an isocurvature perturbation is established:  the energy density in
the scalar field is compensated by a deficit in radiation.  However, the
contraction of the scalar field configuration leads to a clumping of gradient
and kinetic energy at the center of the texture (Fig. 5).  This, in turn,
provides
seed perturbations which cause dark matter and radiation to collapse in a
spherical manner.

\midinsert \vskip 5.5cm
\hsize=6in \raggedbottom
\noindent {\bf Figure 5}: A sketch of the density perturbation produced
by a collapsing texture.  The left graph shows the time evolution of
the field $\chi (r)$ as a function of radius $r$ and time (see
(3.10)).  The contraction of $\chi (r)$ leads to a spatial gradient
energy perturbation at the center of the texture, as illustrated on
the right.  The energy is denoted by $\rho$.  Solid lines denote the
initial time, dashed lines are at time $t + \Delta t$, and dotted
lines correspond to time $t + 2 \Delta t$, where $\Delta t$ is a
fraction of the Hubble expansion time (the typical time scale for the
dynamics).
\endinsert

In both cosmic string and texture models, the fluctuations are non-Gaussian,
which means that the Fourier modes of the density perturbation $\delta \rho$
have nonrandom phases.  Most inflationary Universe models, in contrast,
predict (in linear theory) random phase fluctuations which can be
viewed as a superposition of small amplitude plane wave perturbations with
uncorrelated phases (for some subtle issues see Refs. 12 and 13).
\par
Before discussing some key observations which will allow us to distinguish
between the different models, I will discuss the role of dark matter.  The key
issue is free streaming.  Recall that cold dark matter consists of particles
which have negligible velocity $v$ at $t_{eq}$, the time when sub-horizon
scale perturbations can start growing:
$$
v (t_{eq}) \ll 1 \qquad {\rm (CDM)} \, . \eqno\eq
$$
For hot dark matter, on the other hand:
$$
v (t_{eq}) \sim 1 \qquad {\rm (HDM)} \, . \eqno\eq
$$
Due to their large thermal velocities, it is not possible to establish
HDM perturbations at early times on small scales.  Fluctuations are
erased by free streaming on all scales smaller than the free streaming
length
$$
\lambda_J^c (t) = v (t) z (t) t \eqno\eq
$$
(in comoving coordinates).  For $t > t_{eq}$, the free streaming
length decreases as $t^{-1/3}$.  The maximal streaming length is
$$
\lambda_J^{\rm max} = \lambda^c_J (t_{eq}) \eqno\eq
$$
which for $v (t_{eq}) \sim 0.1$ (appropriate for 25 eV neutrinos)
exceeds the scale of galaxies.
\par
In inflationary Universe models and in the texture theory, the density
perturbations essentially are dark matter fluctuations.  The above free
streaming analysis then shows that, if the dark matter is hot, then
no perturbations on the scale of galaxies
will survive independent of larger-scale structures.  Hence, these
theories are acceptable only if the dark matter is cold.
\par
Cosmic string theories, in contrast, work well - if not even better -
with hot dark matter$^{5,35,38)}$.  The cosmic string seeds survive free
streaming.  The growth of perturbations on small scales $\lambda$ is
delayed (it starts once $\lambda = \lambda_J (t)$) but not
prevented.
\par
Let us summarize the main characteristics of the cosmic string, global
texture, and inflationary Universe theories of structure formation.
Inflation predicts random phase perturbations.  The density peaks will
typically be spherical, and the model is consistent with basic
observations only for CDM.  The global texture and cosmic string
models both give non-random phase perturbations.  The topology is
dominated by spherical peaks for textures, whereas it is planar or
filamentary for
cosmic strings (depending on the small scale structure on the
strings).  Textures requrie CDM, whereas cosmic strings work
better with HDM.

\chapter{Signatures for Cosmics Strings and Textures}
\section{Large-Scale Structure Signatures}
\par
The \underbar{genus curve}$^{39)}$ is a statistical measure for the topology of
large-scale structure.  Given a smooth density field $\rho
(\undertext{x})$, we pick a density $\rho_0$ and consider the surface
$S_{\rho_0}$ where $\rho (\undertext{x}) = \rho_0$ and calculate the
genus $\nu (S)$ of this surface
$$
\nu =  {\rm \# \> of \> holes \> of \>} S - {\rm \# of \> disconnected
\> components \> of \>} S \, . \eqno\eq
$$
The genus curve is the graph of $\nu$ as a function of $\rho_0$.
\par
For perturbations with Poisson statistics, the genus curve can be
calculated analytically (Fig. 6).  The inflationary CDM model in the
linear regime falls in this category.  The genus curve is symmetric
about the mean density $\bar \rho$.  In the texture model, the
symmetry about $\bar \rho$ is broken and the genus curve is shifted to
the left$^{40)}$.  In the cosmic string model, there is a pronounced
asymmetry between $\rho > \bar \rho$ and $\rho < \bar \rho$.  At small
densities, the genus curve measures the (small number) of large voids,
whereas for $\rho > \bar \rho$ the curve picks$^{41)}$ out the large
number of high density peaks which result as a consequence of the
fragmentation of the wakes (Fig. 6).

\midinsert \vskip 10cm
\hsize=6in \raggedbottom
\noindent{\bf Figure 6.} The genus curve of the smoothed mass density
field in a cosmic string wake toy model compared to the symmetric
curve which results in the case of a model with a random distribution
of mass points. The vertical axis is the genus (with genus zero at the
height of the ``x"), the horizontal axis is a measure of density (``0"
denotes average density).
\endinsert

The \undertext{counts in cell statistic}$^{42)}$ can be successfully
applied to distinguish between distributions of galaxies with the same
power spectrum but with different phases.  The statistic is obtained
by dividing the sample volume into equal size cells, counting the
number $f(n)$ of cells containing $n$ galaxies, and plotting $f(n)$ as
a function of $n$.

We$^{43)}$ have applied this statistic to a set of toy models of
large-scale structure.  In each case, the sample volume was (150
Mpc)$^3$, the cell size (3.75 Mpc)$^3$, and the samples contained
90,000 galaxies.  We compared a texture model (galaxies distributed in
spherical clumps separated by 30 Mpc with a Gaussian radial density
field of width 9 Mpc), a cosmic string model dominated by filaments
(all galaxies randomly distributed in filaments of dimensions (60
$\times$ 4 $\times$ 4) Mpc$^3$ with mean separation 30 Mpc, a cosmic
string wake model (same separation and wake dimensions $(40 \times 40
\times 2)$ Mpc$^3$), a cold dark matter model (obtained by Fourier
transforming the CDM power spectrum and assigning random phases), and
a Poisson distribution of galaxies.

\midinsert \vskip 17cm
\hsize=6in \raggedbottom
\noindent{\bf Figure 7.}
The three dimensional counts in cell statistics for a Poisson model
(G), a cold dark matter model (CDM), cosmic string wakes (SW), string
filaments (SF) and textures (T).
\endinsert

As shown in Fig. 7, the predicted curves differ significantly,
demonstrating that this statistic is an excellent one at
distinguishing different theories with the same power spectrum.  The
counts in cell statistic can also be applied to effectively two
dimensional surveys such as single slices of the CFA redshift
survey$^{44)}$.  The predictions of our theoretical toy models are
shown in Fig. 8.

\midinsert \vskip 17cm
\hsize=6in \raggedbottom
\noindent{\bf Figure 8.}
The two dimensional counts in cell statistic for a slice of the
Universe of the dimensions of a CFA slice, evaluated for the same
models as in Fig. 7.
\endinsert

A third statistic which has proved useful$^{45)}$ in distinguishing
models with Gaussian and non-Gaussian phases is the void probability
function $p(R)$, the probability that a sphere of radius $R$ contains
no galaxies.

\subsection{Signatures in the Microwave Background}

Inflationary Universe models predict essentially random phase
fluctuations in the microwave background with a scale invariant
spectrum $(n = 1)$.  Small deviations from scale invariance are model
dependent and were discussed in detail by D. Salopek at this
meeting$^{46)}$.  In all models, the amplitude must be consistent with
structure formation.  As mentioned in Section 2, the COBE
discovery$^{1)}$ of anisotropies in the CMB has provided severe
constraints on inflationary models.  They are only consistent with the
present data if the bias parameter $b$ is about 1, which must be
compared to the value $b \simeq 2.5$ which is the best value for
galaxy formation in this model$^{17)}$.  Note that full sky coverage
is not essential for testing inflationary models since in any set of
local observations of $\delta T/T$, the results will form a Gaussian
distribution about the r.m.s. value.

\midinsert \vskip 7.5cm
\hsize=6in \raggedbottom
\noindent{\bf Figure 9.}
Sketch of the mechanism producing linear discontinuities in the
microwave temperature for photons $\gamma$ passing on different sides
of a moving string $S$ (velocity $v$).  ${\cal O}$ is the observer.
Space perpendicular to the string is conical (deficit angle $\alpha$).
\endinsert

Cosmic string models predict non-Gaussian temperature anisotropies.
One mechanism gives rise to localized linear temperature
discontinuities$^{47)}$; its origin is illustrated in Fig. 9.  Photons
passing on different sides of a long straight string moving with
velocity $v$ reach the observer with a Doppler shift
$$
{\delta T\over T} \sim 8 \pi G \mu v \gamma (v) \, . \eqno\eq
$$
To detect such discontinuities, an appropriate survey strategy (
e.g. full sky survey) with small angular resolution is crucial.  The
distribution of strings also gives rise to Sachs-Wolfe type
anisotropies.$^{48)}$

The theoretical error bars in the normalization of CMB anisotropies
from strings are rather large -- a direct consequence of the fact that
the precise form of the scaling solution for the string network is not
well determined.  Nevertheless, we can consider a fixed set of cosmic
string parameters and ask whether the normalizations of $G\mu$ from
large-scale structure data and from COBE are consistent.  This has
been done numerically in Ref. 2, and using an analytical toy model in
Ref. 3.

The analytical model$^{3)}$ is based on adding up as a random walk the
individual Doppler shifts from strings which the microwave photons
separated by angular scale $\theta$ pass on different sides, and
using this method to compute $\Delta T/T (\theta)$.  Using the
Bennett-Bouchet$^{27)}$ string parameters, the result for $G \mu$
becomes
$$
G\mu = (1.3 \pm 0.5) 10^{-6} \, ,  \eqno\eq
$$
in good agreement with the requirements from large-scale structure
formation$^{5,15)}$.

To detect the predicted anisotropies from textures, it is again
essential to have a full sky survey.  However, large angular
resolution is adequate this time, since the specific signature for
textures is a small number $( \sim 10)$ of hot and cold disks with
amplitude$^{18)}$
$$
{\delta T\over T} \sim 0.06 \times 16 \> \pi \> G \eta^2 \sim 3 \cdot 10^{-5}
\eqno\eq
$$
and angular size of about $10^\circ$.
The hot and cold spots are due to photons
falling into the expanding Goldstone boson radiation field which
results after texture collapse or due to photons climbing out of the
potential well of the collapsing texture$^{49)}$ (see Fig. 10).

\midinsert \vskip 5cm
\hsize=6in \raggedbottom
\noindent{\bf Figure 10.} Space-time diagram of a collapsing texture
(backward light cone) and the resulting expanding Goldstone boson
radiation (forward light cone). Unwinding of the texture occurs at
point ``TX". The light ray $\gamma_2$ falls into the potential well
and is blueshifted, the ray $\gamma_1$ is redshifted.
\endinsert

Note that the texture model is not ruled out by the recent COBE
results.  The amplitude (5.4) is lower than the pixel sensitivity of
the COBE maps.  However, the predicted quadrupole CMB anisotropy
(normalizing $\eta$ by the large-scale structure data) exceeds the
COBE data print by a factor of between 2.5 and 4$^{19,20)}$.  Hence,
biasing must be invoked in order to try to explain the large-scale
structure data given the reduced value of $\eta$ mandated by the
discovery of CMB anisotropies.

\chapter{Conclusions}

Topological defect models of structure formation generically give rise
to a scale invariant power spectrum and are hence in good agreement
with the recent COBE results on anisotropies of the CMB.  The
amplitude of the quadrupole temperature fluctuation can be used to
normalize the models.  For cosmic strings, the resulting normalization
agrees well with the normalization from large-scale structure data.
For textures, there is a mismatch which requires introducing biasing.
For textures, the situation is comparable to that in the CDM model,
where COBE demands a bias parameter $b \simeq 1$, whereas galaxy
formation is said to demand$^{17)}$ $b \simeq 2.5$.

It was emphasized that r.m.s. data intrinsically is unable to
differentiate between topological defect models (with non-random
phases) and inflationary models (with random phases).  We need
statistics which are sensitive to nonrandom phases.  We propose using
two dimensional analogs of the statistics discussed in Section 5, {\it
e.g.} the counts in cell statistic (see also Ref. 50).

The most economical model for structure formation may be the model
based on cosmic strings and hot dark matter.  It requires no new
particles (although it does require a finite neutrino mass), it agrees
well with COBE and with the CFA redshift data, and it has clear
signatures both for large-scale structure and CMB statistics.

\ack
\par
I wish to thank Norma Sanchez for organizing this school and for the
invitation to lecture.  This work has been supported in part by the US
Department of Energy under Grant DE-AC02-03130 Task A.

\endpage

\REF\one{G. Smoot et al., {\it Astrophys. J. (Lett.)}, in press
(1992) [COBE preprint 92-04].}
\REF\two{D. Bennett, A. Stebbins and F. Bouchet,`The Implications of
the COBE-DMR Results for Cosmic Strings' Fermilab preprint
92/144-A (1992).}
\REF\three{L. Perivolaropoulos, `COBE vs. Cosmic Strings: An
Analytical Model' CFA preprint 3467 (1992).}
\REF\four{G. Blumenthal, S. Faber, J. Primack and M. Rees, {\it
Nature} {\bf 311}, 517 (1984);\nextline
M. Davis, G. Efstathiou, C. Frenk and S. White, {\it Ap. J.} {\bf
292}, 371 (1985).}
\REF\five{R. Brandenberger, N. Kaiser, D. Schramm and N. Turok, {\it
Phys. Rev. Lett.} {\bf 59}, 2371 (1987);\nextline
R. Brandenberger, L. Perivolaropoulos and A. Stebbins, {\it Int. J.
Mod. Phys.} {\bf 5}, 1633 (1990).}
\REF\six{N. Turok, {\it Phys. Rev. Lett.} {\bf 63}, 2625 (1989).}
\REF\seven{Ya B. Zel'dovich, {\it Mon. Not. R. astron. Soc.} {\bf
192}, 663 (1980); \nextline
A. Vilenkin, {\it Phys. Rev. Lett.} {\bf 46}, 1169 (1981).}
\REF\eight{M. Barriola and A. Vilenkin, {\it Phys. Rev. Lett.} {\bf
63}, 341 (1989).}
\REF\nine{S. Rhie and D. Bennett, {\it Phys. Rev. Lett.} {\bf 65},
1709 (1991).}
\REF\ten{R. Davis, {\it Phys. Rev.} {\bf D35}, 3705 (1987).}
\REF\eleven{C. Hill, D. Schramm and J. Fry, {\it Comm. Nucl. Part.
Phys.} {\bf 19}, 25 (1989).}
\REF\twelve{L. Grishchuk and Y. Sidorov, {\it Class. Quant. Grav.}
{\bf 6}, L161 (1989);\nextline
L. Grishchuk and Y. Sidorov, {\it Phys. Rev.} {\bf D42}, 3413
(1990);\nextline
L. Grishchuk, `Quantum Mechanics of the Primordial Cosmological
Perturbations', to appear in the proceedings of the Sixth Marcel
Grossmann Meeting (Kyoto, 1991).}
\REF\thirteen{T. Prokopec, `Entropy of the Squeezed Vacuum', Brown
Univ. preprint BROWN-HET-861 (1992).}
\REF\fourteen{A. Guth and S.-Y. Pi, {\it Phys. Rev. Lett.} {\bf 49},
1110 (1982); \nextline
S. Hawking, {\it Phys. Lett.} {\bf 115B}, 295 (1982); \nextline
A. Starobinsky, {\it Phys. Lett.} {\bf 117B}, 175 (1982); \nextline
J. Bardeen, P. Steinhardt and M. Turner, {\it Phys. Rev.} {\bf D28},
679 (1983); \nextline
R. Brandenberger and R. Kahn, {\it Phys. Rev.} {\bf D29}, 2172
(1984).}
\REF\fifteen{N. Turok and R. Brandenberger, {\it Phys. Rev.} {\bf
D33}, 2175 (1986); \nextline
A. Stebbins, {\it Ap. J (Lett.)} {\bf 303}, L21 (1986); \nextline
H. Sato, {\it Prog. Theor. Phys.} {\bf 75}, 1342 (1986).}
\REF\sixteen{E. Wright et al., COBE preprint 92-06 (1992).}
\REF\seventeen{S. White, C. Frenk, M. Davis and G. Efstathiou, {\it
Ap. J.} {\bf 313}, 505 (1987).}
\REF\eighteen{R. Durrer and D. Spergel, `Microwave Anisotropies from
Texture Seeded Structure Formation', Princeton Univ. preprint PUPT-91-1247
(1991).}
\REF\nineteen{D. Bennett and S. Rhie, `COBE's Constraints on the
Global Monopole and Texture Theories of Cosmic Structure Formation',
Lawrence Livermore preprint UCRL-JC-111244 (1992).}
\REF\twenty{D. Spergel and N. Turok, private communication (1992).}
\REF\twentyone{T.W.B. Kibble, {\it J. Phys.} {\bf A9}, 1387 (1976).}
\REF\twentytwo{H. Nielsen and P. Olesen, {\it Nucl. Phys.} {\bf B61}, 45
(1973).}
\REF\twentythree{T. Prokopec, {\it Phys. Lett.} {\bf 262B}, 215 (1991).}
\REF\twentyfour{E.P.S. Shellard, {\it Nucl. Phys.} {\bf B283}, 624 (1987).}
\REF\twentyfive{T. Vachaspati and A. Vilenkin, {\it Phys. Rev.} {\bf D31},
3052 (1985).}
\REF\twentysix{A. Vilenkin, {\it Phys. Rep.} {\bf 121}, 263 (1985);
\nextline
N. Turok, `Phase Transitions as the Origin of Large-Scale Structure',
in `Particles, Strings and Supernovae' (TASI88) ed. by A. Jevicki and
C.-I. Tan (World Scientific, Singapore, 1989); \nextline
R. Brandenberger, {\it J. Phys.} {\bf G15}, 1 (1989).}
\REF\twentyseven{A. Albrecht and N. Turok, {\it Phys. Rev. Lett.} {\bf 54},
1868 (1985); \nextline
D. Bennett and F. Bouchet, {\it Phys. Rev. Lett.} {\bf 60}, 257
(1988);\nextline
B. Allen and E.P.S. Shellard, {\it Phys. Rev. Lett.} {\bf 64}, 119
(1990);\nextline
A. Albrecht and N. Turok, {\it Phys. Rev.} {\bf D40}, 973 (1989).}
\REF\twentyeight{N. Turok, {\it Phys. Scripta} {\bf T36}, 135 (1991).}
\REF\twentynine{R. Leese and T. Prokopec,
{\it Phys. Rev.} {\bf D44}, 3749 (1991).}
\REF\thirty{T. Prokopec, A. Sornborger and R. Brandenberger,
{\it Phys. Rev.} {\bf D45}, 1971 (1992).}
\REF\thirtyone{R. Brandenberger, in `Physics of the Early Universe',
Scottish Univ. Summer School in Physics 36, 1989, ed. by J. Peacock,
A. Heavens and A. Davies (IOP Publishing, Bristol, 1990).}
\REF\thirtytwo{A. Vilenkin, {\it Phys. Rev.} {\bf D23}, 852 (1981).}
\REF\thirtythree{J. Silk and A. Vilenkin, {\it Phys. Rev. Lett.} {\bf 53},
1700 (1984).}
\REF\thirtyfour{T. Vachaspati, {\it Phys. Rev. Lett.} {\bf 57}, 1655
(1986); \nextline
A. Stebbins, S. Veeraraghavan, R. Brandenberger, J. Silk and N. Turok,
{\it Ap. J.} {\bf 322}, 1 (1987).}
\REF\thirtyfive{L. Perivolaropoulos, R. Brandenberger and A. Stebbins, {\it
Phys
Rev.} {\bf D41}, 1764 (1990).}
\REF\thirtysix{B. Carter, {\it Phys. Rev.} {\bf D41}, 3869 (1990).}
\REF\thirtyseven{D. Vollick, {\it Phys. Rev.} {\bf D45}, 1884 (1992);
\nextline
T. Vachaspati and A. Vilenkin, {\it Phys. Rev. Lett.} {\bf 67}, 1057
(1991).}
\REF\thirtyeight{R. Brandenberger, N. Kaiser and N. Turok, {\it Phys. Rev.}
{\bf D36}, 2242 (1987).}
\REF\thirtynine{J. Gott, A. Melott and M. Dickinson, {\it Ap. J.} {\bf
306}, 341 (1986).}
\REF\forty{A. Gooding, C. Park, D. Spergel, N. Turok and J. Gott,
{\it Ap. J.} {\bf 393}, 42 (1992).}
\REF\fortyone{J. Gerber and R. Brandenberger, `Topology of Large-Scale
Structure in a Cosmic String Wake Model', Brown Univ. preprint
BROWN-HET-829 (1991).}
\REF\fortytwo{W. Saslaw, {\it Ap. J.} {\bf 297}, 49 (1985).}
\REF\fortythree{S. Ramsey, Senior thesis, Brown Univ. (1992);
\nextline
S. Ramsey and R. Brandenberger, in preparation (1992).}
\REF\fortyfour{V. de Lapparent, M. Geller and J. Huchra, {\it Ap. J.
(Lett.)} {\bf 302}, L1 (1986).}
\REF\fortyfive{E. Valentini and R. Brandenberger, in preparation
(1992); \nextline
D. Weinberg and S. Cole, `Non-Gaussian Fluctuations and the Statistics
of Galaxy Clustering', Berkeley preprint CfPA-TH-91-025 (1991).}
\REF\fortysix{D. Salopek, these proceedings.}
\REF\fortyseven{N. Kaiser and A. Stebbins, {\it Nature} {\bf 310},
391 (1984).}
\REF\fortyeight{J. Traschen, N. Turok and R. Brandenberger, {\it Phys.
Rev.} {\bf D34}, 919 (1986); \nextline
S. Veeraraghavan and A. Stebbins, `Large-Scale Microwave Anisotropy
from Gravity Seeds', Fermilab preprint 92/147-A (1992).}
\REF\fortynine{N. Turok and D. Spergel, {\it Phys. Rev. Lett.} {\bf
64}, 2736 (1990).}
\REF\fifty{P. Coles, {\it Mon. Not. R. astron. Soc.} {\bf 234}, 509
(1988); \nextline
P. Coles and M. Plionis, {\it Mon. Not. R. astron. Soc.} {\bf 250}, 75
(1991).}
\refout

\bye